\documentclass[prl,twocolumn,showpacs,preprintnumbers,amsmath,amssymb]{revtex4}

\usepackage{graphicx}
\usepackage{dcolumn,color}
\usepackage{bm}

\begin{document}


\boldmath
\title{First observation of in-medium modifications of the $\omega$ meson}
\unboldmath

\author{
  D.~Trnka~$^1$,
  G.~Anton~$^2$,
  J.~C.~S.~Bacelar~$^3$,
  O.~Bartholomy~$^4$,
  D.~Bayadilov~$^{4,~8}$,
  Y.A.~Beloglazov~$^8$,
  R.~Bogend\"orfer~$^2$,
  R.~Castelijns~$^3$,
  V.~Crede~$^{4,^\ast}$,
  H.~Dutz~$^5$,
  A.~Ehmanns~$^4$,
  D.~Elsner~$^5$,
  R.~Ewald~$^5$,
  I.~Fabry~$^4$,
  M.~Fuchs~$^4$,
  K.~Essig~$^4$,
  Ch.~Funke~$^4$,
  R.~Gothe~$^{5,^\diamond}$,
  R.~Gregor~$^1$,
  A.~B.~Gridnev~$^8$,
  E.~Gutz~$^4$,
  S.~H\"offgen~$^5$,
  P.~Hoffmeister~$^4$,
  I.~Horn~$^4$,
  J.~H\"ossl~$^2$,
  I.~Jaegle~$^7$,
  J.~Junkersfeld~$^4$,
  H.~Kalinowsky~$^4$,
  Frank~Klein~$^5$,
  Fritz~Klein~$^5$,
  E.~Klempt~$^2$,
  M.~Konrad~$^5$,
  B.~Kopf~$^{6,9}$,
  M.~Kotulla~$^7$,
  B.~Krusche~$^7$,
  J.~Langheinrich~$^{5,^\diamond}$,
  H.~L\"ohner~$^3$,
  I.V.~Lopatin~$^8$,
  J.~Lotz~$^4$,
  S.~Lugert~$^1$,
  D.~Menze~$^5$,
  J.~G.~Messchendorp~$^3$,
  T.~Mertens~$^7$,
  V.~Metag~$^1$,
  C.~Morales~$^5$,
  M.~Nanova~$^1$,
  R.~Novotny~$^1$,
  M.~Ostrick~$^5$,
  L.~M.~Pant~$^{1,^\dagger}$,
  H.~van Pee~$^1$,
  M.~Pfeiffer~$^1$,
  A.~Roy~$^{1,^\ddagger}$,
  A.~Radkov~$^8$,
  S.~Schadmand~$^{1,^\star}$,
  Ch.~Schmidt~$^4$,
  H.~Schmieden~$^5$,
  B.~Schoch~$^5$,
  S.~Shende~$^3$,
  G.~Suft~$^2$,
  V.~V.~Sumachev~$^8$,
  T.~Szczepanek~$^4$,
  A.~S\"ule~$^5$,
  U.~Thoma~$^{1,4}$,
  R.~Varma~$^{1,^\ddagger}$,
  D.~Walther~$^5$,
  Ch.~Weinheimer~$^{4,^+}$,
  Ch.~Wendel~$^4$\\
(The CBELSA/TAPS Collaboration)
}
\affiliation{
  $^1$II. Physikalisches Institut, Universit\"at Gie{\ss}en, Germany\\
  $^2$Physikalisches Institut, Universit\"at Erlangen, Germany\\
  $^3$KVI, Groningen, The Netherlands\\
  $^4$\mbox{Helmholtz-Institut f\"ur Strahlen- u. Kernphysik, Universit\"at Bonn, Germany}\\
  $^5$Physikalisches Institut, Universit\"at Bonn, Germany\\
  $^6$\mbox{Institut f\"ur Kern- und Teilchenphysik, TU Dresden, Germany}\\
  $^7$Physikalisches Institut, Universit\"at Basel, Switzerland\\
  $^{8}$Petersburg Nuclear Physics Institute, Gatchina, Russia\\
  $^9$Physikalisches Institut, Universit\"at Bochum, Germany\\
  $^{\diamond}$ now at University of South Carolina, Columbia, USA\\
  $^\ast$ now at Florida State University, USA\\
  $^\star$\mbox{now at Institut f\"ur Kernphysik, FZ J\"ulich, Germany}\\
  $^+$\mbox{now at Institut f\"ur Kernphysik, Universit\"at M\"unster, Germany}\\
  $^\dagger$ \mbox{on leave from Nuclear Physics Division, BARC, Mumbai, India} \\
  $^\ddagger$ \mbox{on leave from Department of Physics, I.I.T. Powai, Mumbai, India}\\
  }%
\date{\today}

\begin{abstract}
The photoproduction of $\omega$ mesons on nuclei has been
investigated using the Crystal Barrel/TAPS experiment
at the ELSA tagged photon facility in Bonn. The aim is to study possible
in-medium modifications of the $\omega$ meson via the reaction 
$\gamma + A \rightarrow \omega + X \rightarrow \pi^0 \gamma + X^\prime$.
Results obtained for Nb are compared to a reference measurement on a $\rm{LH_2}$ target.
While for recoiling, long-lived mesons ($\pi^0$, $\eta$ and $\eta^\prime$), which decay outside 
of the nucleus,
a difference in the lineshape for the two data samples is not observed,
we find a significant enhancement 
towards lower masses for $\omega$ mesons produced on the Nb target.
For momenta less than $\rm{500~MeV/c}$ an in-medium $\omega$ meson mass
of 
$\rm{M_{medium}=[722_{-2}^{+2}(stat)_{-5}^{+35}(syst)]}~\rm{MeV/c^2}$ has been deduced
at an estimated average nuclear
density of $0.6~\rho_0$.
\end{abstract}

\boldmath
\maketitle
\unboldmath
The modification of experimentally observable properties of vector
mesons such as mass and width, when embedded in a dense medium, 
is one of the most fundamental research issues in hadron physics.
While for
composite systems like molecules, atoms or nuclei,
the mass of the system is almost completely 
(apart from small binding energy effects) governed by the sum of the masses of the constituents,
this is no longer true in the hadronic sector.
Here, the hadron masses are much larger than the summed masses of the constituents,
the u-, d- and s-quarks.
One possible
interpretation is that the masses of hadrons are generated
dynamically \cite{wilczek}.
Furthermore, hadron masses can be associated with the
spontaneous breaking of chiral symmetry. 
In nuclear matter (and at high temperatures) this symmetry is
predicted to be at least partially restored. As a consequence, the
properties of hadrons are expected to be modified 
(see e.g. \cite{meis,klimt,brown,hats,shu1}).
\par
A variety of theoretical models predict a
lowering of the in-medium mass of vector mesons even at
normal nuclear matter density $\rho_0$. For the $\omega$ meson a drop of the mass by 20 to 150 
$\rm{MeV/c^2}$
and a broadening of the width up to 60 $\rm{MeV/c^2}$ has been predicted
(e.g. \cite{renk,klingl,riek,lyk,weid,eff}).
However, the discussion in the literature is very controversial. 
Even upward mass shifts \cite{zsch} or the appearance of additional peaks \cite{lutz}
have been suggested by some authors. This situation underlines the 
importance of experimental results.
\begin{figure*}
  \hspace*{-0.2cm}
  \includegraphics[width=0.68\columnwidth]{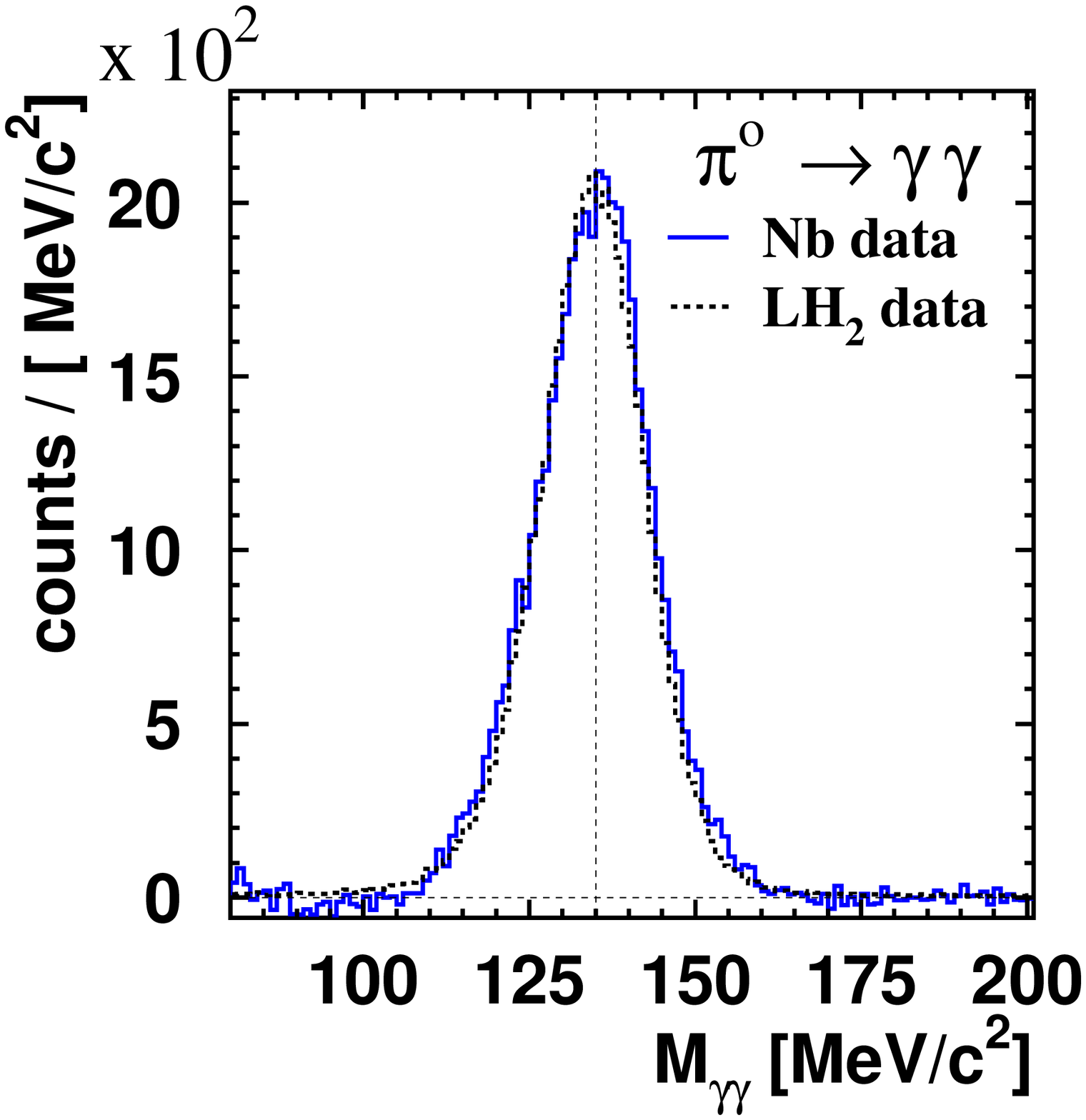}\hspace*{-0.1cm}
  \includegraphics[width=0.68\columnwidth]{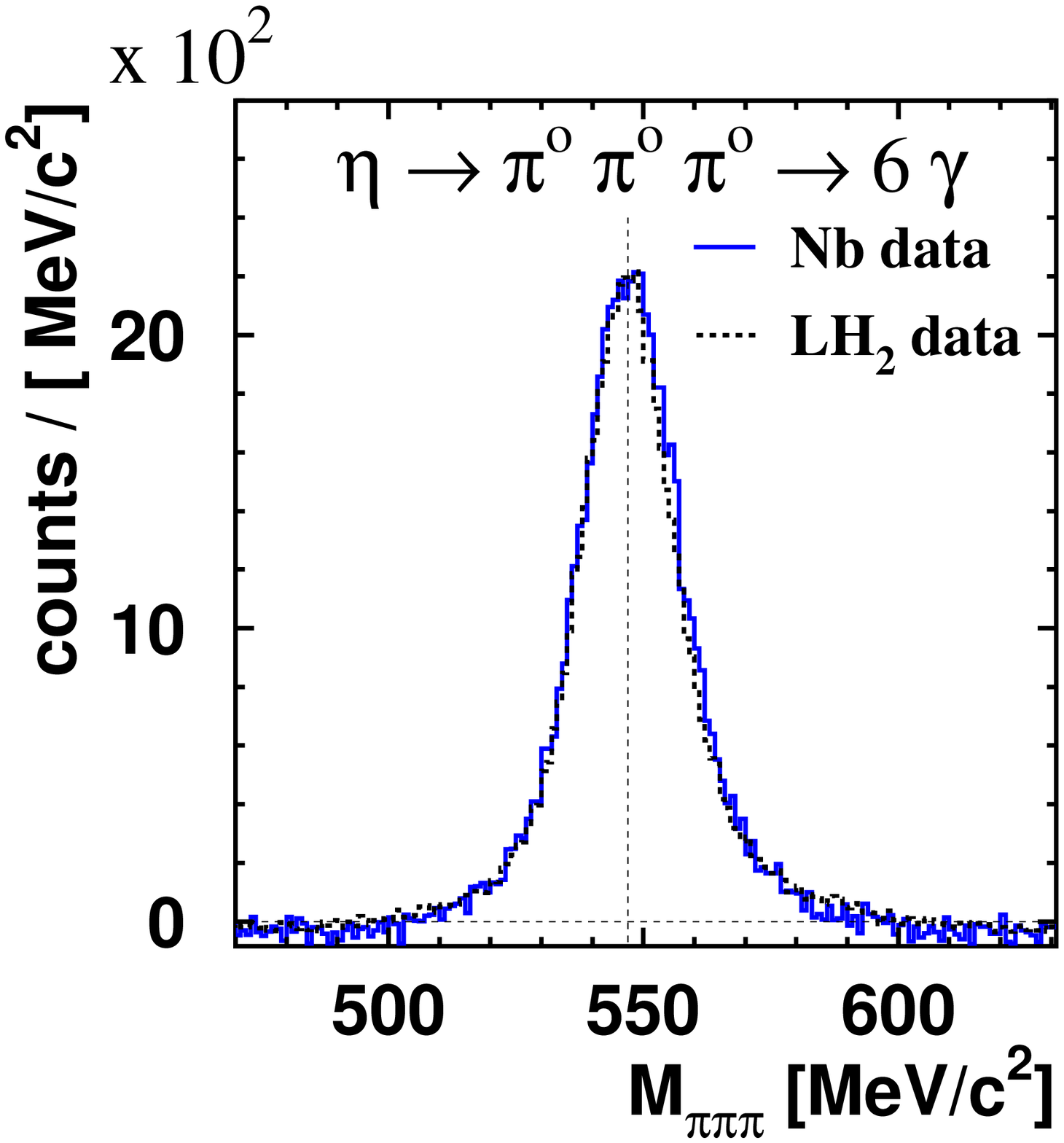}\hspace*{-0.1cm}
  \includegraphics[width=0.68\columnwidth]{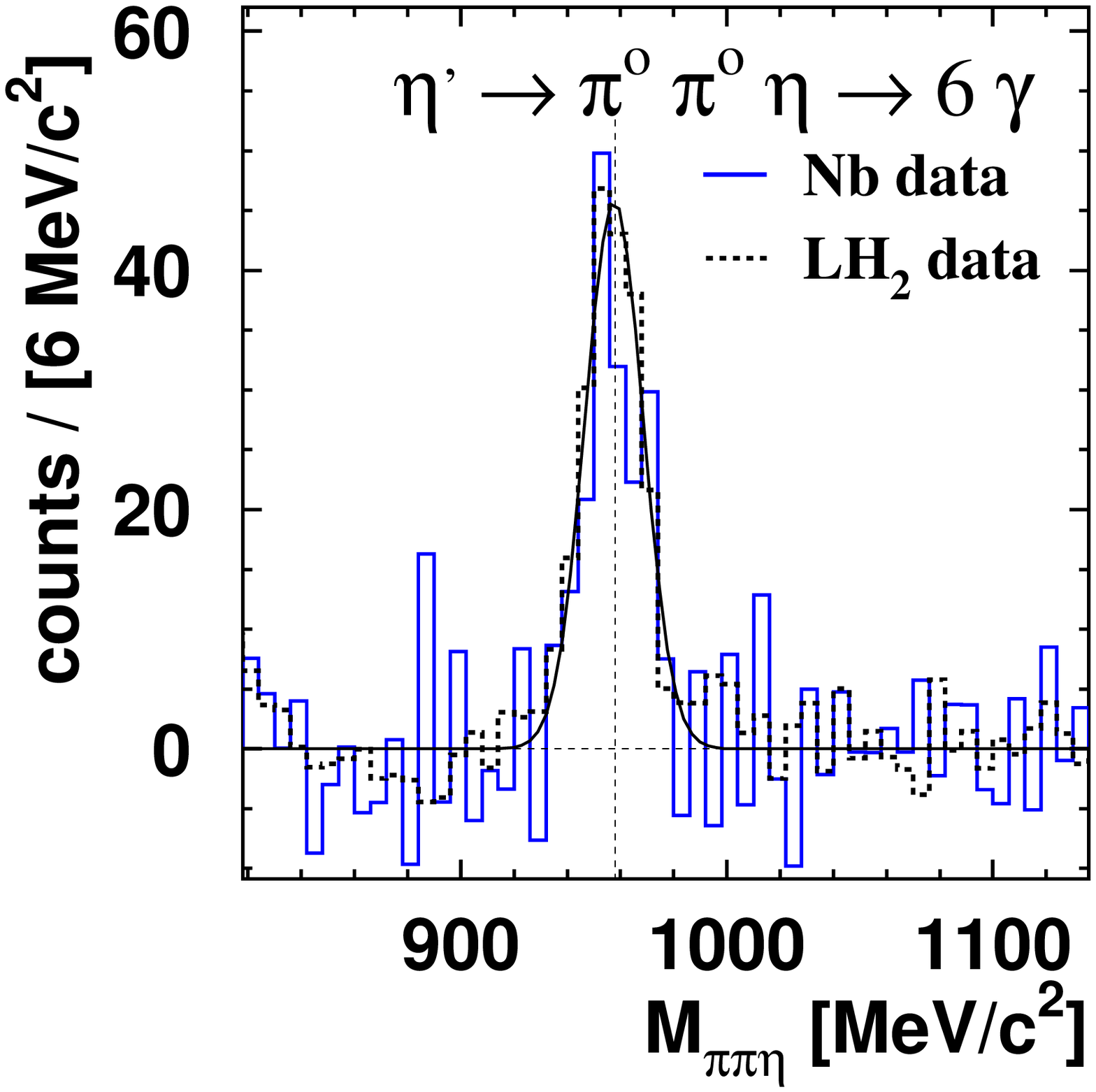}\hspace*{-0.1cm}
   \vspace*{-0.6cm}
  \caption[Longlived mesons]{(color online). Background subtracted invariant mass distributions
 for the long-lived mesons $\pi^0$, $\eta$ and $\eta^\prime$.
The solid  and dashed histograms correspond to data taken with a
Nb and $\rm{LH_2}$ target, respectively.
The meson lineshapes are reproduced by Monte Carlo simulations, as demonstrated for the 
$\eta^\prime$ signal (solid curve).
\vspace*{-0.4cm}
\label{cal_check}}
\end{figure*}
\par
Several previous experiments have studied the properties of vector 
mesons in hot and dense matter.
Dilepton spectroscopy  
allows to measure the in-medium properties without distortion due to 
final state interactions (FSI).
The CERES collaboration at CERN, for example,  
measured the low-mass $e^+e^-$ pair production in heavy-ion 
collisions and observed an enhancement in the mass range
of $0.3~\rm{GeV/c^2} \le m_{e^+e^-} \le 0.7~\rm{GeV/c^2}$  over the yield expected from the known
sources in pp collisions \cite{ag1,ag2}.
More recently, analyzing $\pi^+\pi^-$ pairs the STAR experiment at RHIC observed 
a decrease of the
$\rho$ meson in-medium mass in peripheral
Au+Au collisions \cite{star}.
The KEK-PS E325 collaboration 
investigated $p+A$ reactions at 12 GeV \cite{kek} and reported an enhancement
in the $e^+e^-$ invariant mass spectra  in
the region of $0.6 ~\rm{GeV/c^2} \le m_{e^+e^-} \le 0.77~\rm{GeV/c^2}$.
Presently, an experiment performed at JLAB using a photon beam 
is being analyzed \cite{jlab}.
At GSI, it has been proposed \cite{gsi} to perform pion induced experiments 
with the HADES \cite{gsi1} detector system.
\par
All $e^+e^-$ experiments 
suffer from the small branching ratios (BR) of vector mesons 
into dileptons, which are in the order of $10^{-5} - 10^{-4}$. 
In addition, the comparable $e^+e^-$ decay rates for $\omega$- and $\rho$ mesons make
it difficult 
to isolate an $\omega$ 
signal from the $e^+e^-$ invariant mass spectrum \cite{riek}. 
An alternative and promising approach to investigate in-medium modifications of the
$\omega$ meson is to study the $\omega \rightarrow \pi^0 \gamma$
decay mode,  as pointed out in \cite{muehl,messch,sib}.
An essential advantage of this decay channel is the large BR
of almost $9 \%$, three orders of magnitude larger than the decay into
dileptons.
Furthermore, this
mode is a clean and exclusive probe to study the $\omega$ in-medium
properties since the $\rho \rightarrow \pi^0 \gamma$ BR is only 
$6.8 \cdot 10^{-4}$ and therefore suppressed by two orders of
magnitude relative to the $\omega$ BR into this channel. 
However, the disadvantage is a possible rescattering of the $\pi^0$ within the nuclear medium,
which would distort the deduced $\omega$ invariant mass distribution.
Pion rescattering within the nucleus proceeds 
predominately via the formation of an intermediate $\Delta$ resonance.  
Due to the kinematics of the $\Delta$ resonance decay 
the distorted events are predicted to accumulate at approx. 500 $\rm{MeV/c^2}$, 
far below the nominal $\omega$ invariant mass.
This leads to a small contribution of only about $3 \%$ in the  
mass range of interest, $0.6~\rm{GeV/c^2} < M_{\pi^0 \gamma} < 0.9~\rm{GeV/c^2}$.
Moreover, the authors of \cite{messch} and \cite{muehl}
have demonstrated that the constraint on the kinetic energy $T_{\pi^0} > 150$ MeV suppresses 
the FSI down to the $1 \%$ level. 
\par
Only $\omega$ mesons decaying inside the nucleus carry information on the in-medium
properties. To enhance the in-medium decay probability, the vector meson decay length 
$L_\omega = p_\omega / m_\omega \Gamma_\omega$ should be less than the nuclear radius. 
This can be achieved by applying a kinematic cut on the 3-momentum of the $\omega$
meson.
But still, only a fraction of the $\omega$ mesons will 
decay inside the nucleus. Thus, one expects the $\pi^0 \gamma$ invariant mass spectra
to show a superposition of decays outside of the target at the vacuum mass peak position 
(782 $\rm{MeV/c^2}$)
with modified decays inside the nucleus \cite{messch}.
\begin{figure*}
  \hspace*{0.cm}
  \includegraphics[width=0.71\columnwidth]{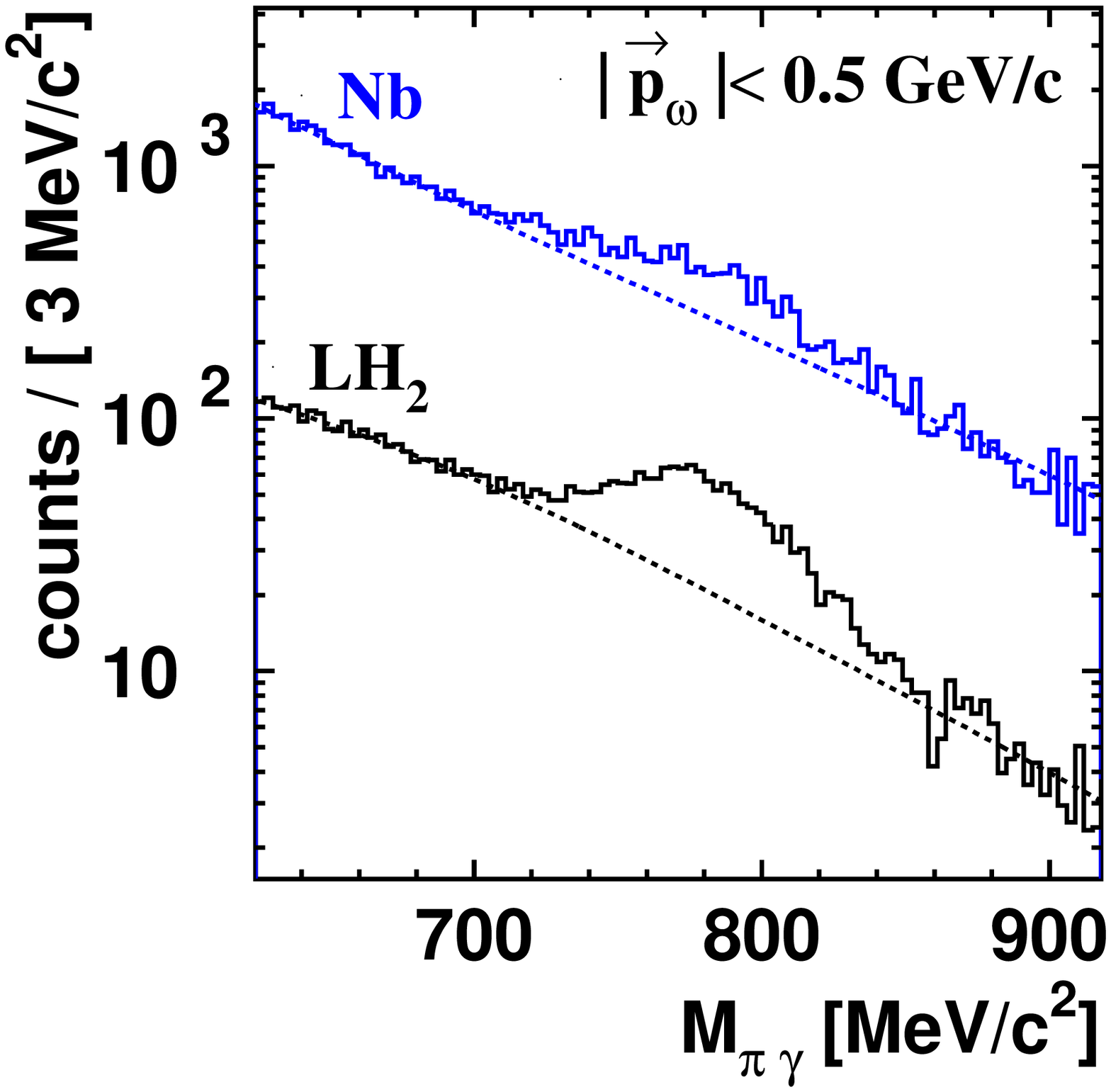}\hspace*{-0.5cm}
  \includegraphics[width=0.71\columnwidth]{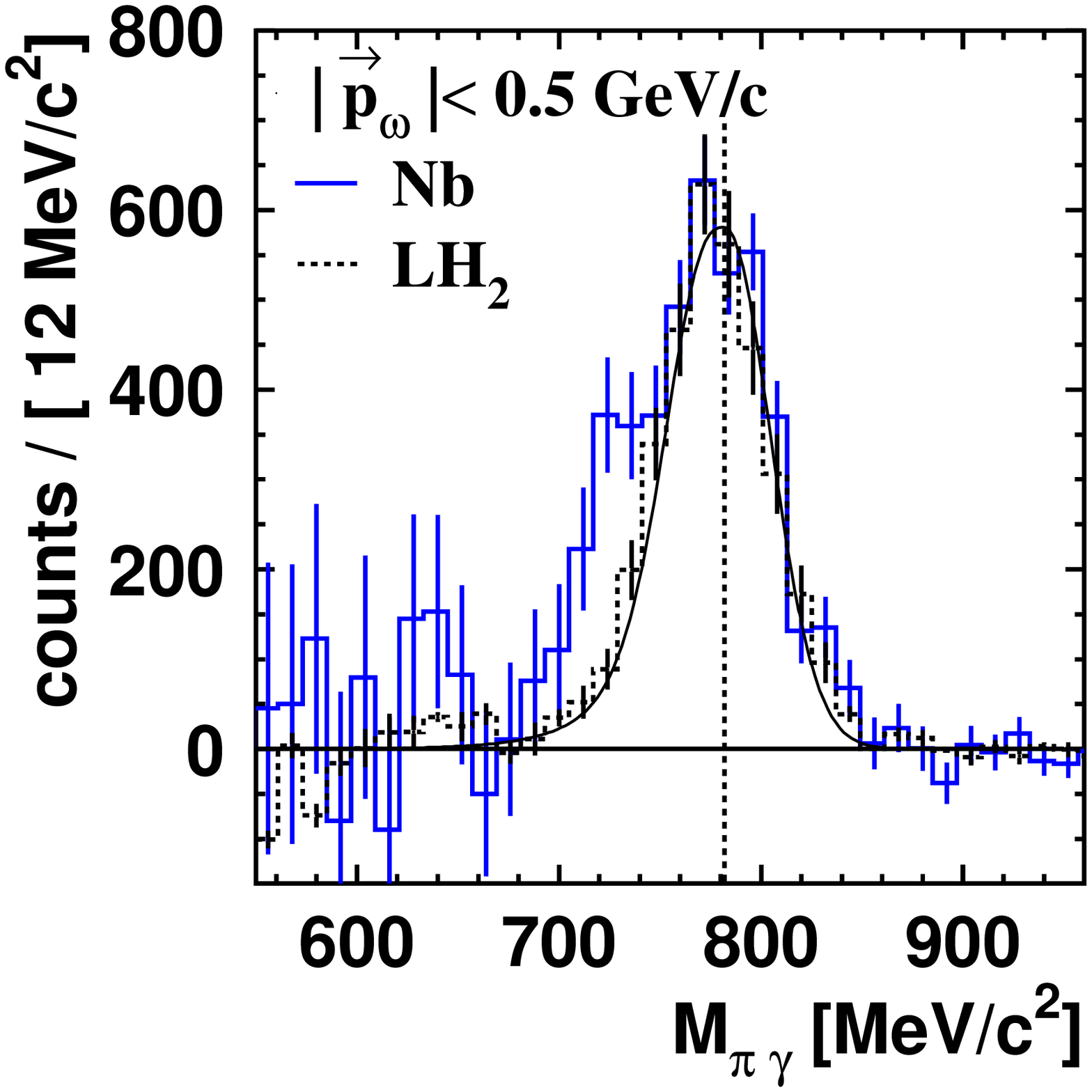}\hspace*{-0.2cm}
  \includegraphics[width=0.71\columnwidth]{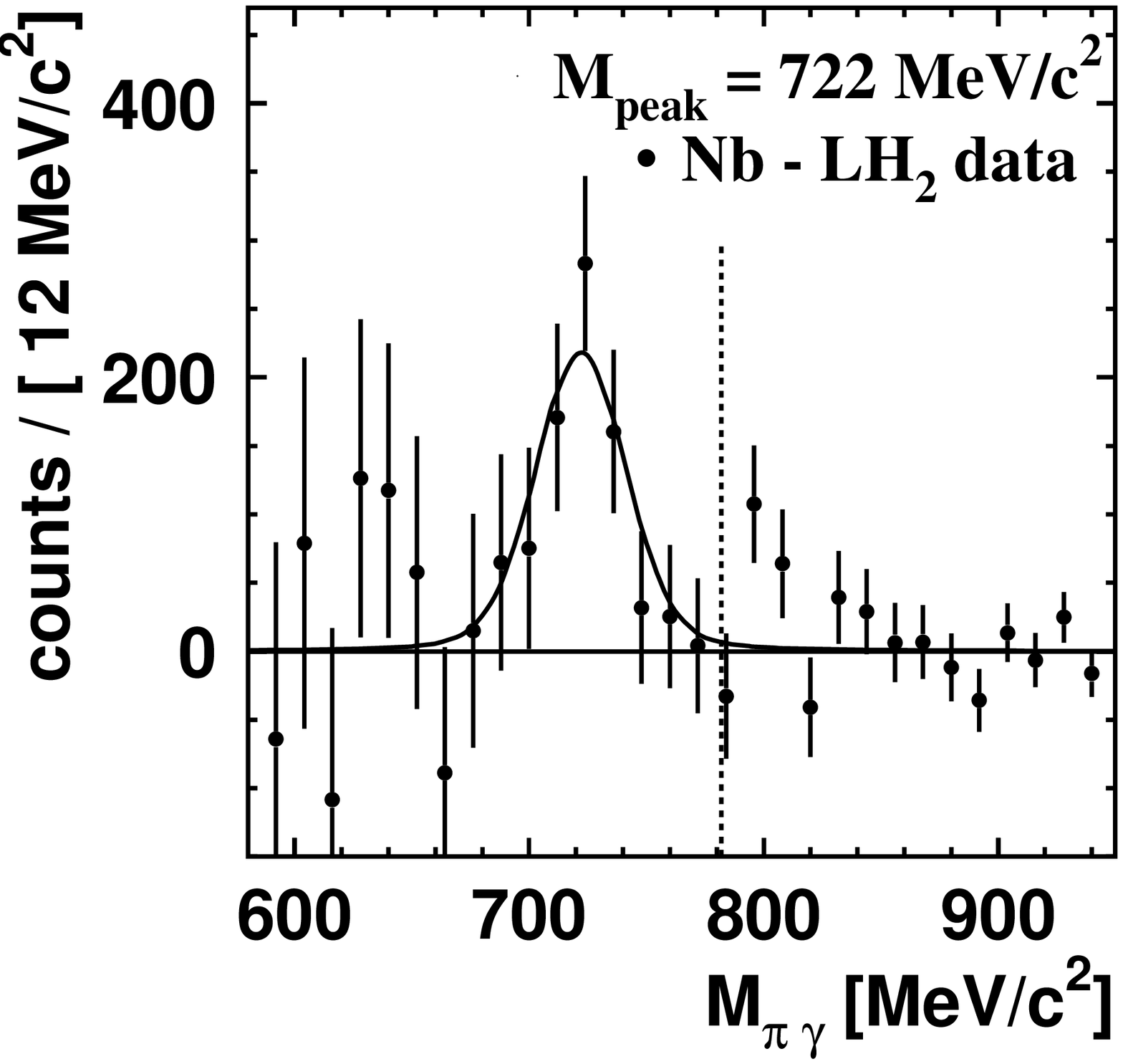}\hspace*{-0.2cm}
  \vspace*{-0.37cm}
  \caption[omega inviariant mass]{(color online). Left panel: Inclusive $\pi^0 \gamma$ invariant mass
spectra for $\omega$ momenta less than 500 $\rm{MeV/c}$.
Upper histogram: Nb data,  
lower histogram: $\rm{LH_2}$ target reference measurement. The dashed lines indicate fits to the
respective background.
Center panel: $\pi^0 \gamma$ invariant mass for the Nb data (solid histogram)
and $\rm{LH_2}$ data (dashed histogram) after background subtraction. The error bars show  
statistical uncertainties only. The solid curve represents the simulated lineshape
for the $\rm{LH_2}$ target.
Right panel: In-medium decays of $\omega$ mesons along with a Voigt fit to the 
data (see text).
The vertical line indicates the vacuum $\omega$ mass of 782 $\rm{MeV/c^2}$.
\label{om_mom}}
\vspace*{-0.47cm}  
\end{figure*}
\par
The experiment was performed at the 
{\bf EL}ectron {\bf S}tretcher {\bf A}ccelerator
(ELSA) in Bonn, using a 2.8 GeV electron beam. The photon beam was produced 
via bremsstrahlung.
A magnetic spectrometer (tagger)
was used to determine the photon beam energies within the tagged photon range of 0.64 to 2.53 GeV.
The Nb and $\rm{LH_2}$ targets had thicknesses of 1 mm and 53~mm, respectively, and 
30 mm in diameter. 
The targets were mounted in the center of the Crystal Barrel detector (CB), a photon
calorimeter consisting of 1290 CsI(Tl) crystals ($\sim$~16~
radiation lengths $X_0$) with an angular coverage of $30^\circ$ up to $168^\circ$ in the polar angle
and a complete azimuthal angle coverage. Inside the CB, covering its full acceptance,
a three-layer scintillating fiber detector
(513 fibers of 2~mm diameter) was installed for charged particle identification.
Reaction products emitted in forward direction
were detected in the TAPS detector. TAPS consisted of
528 hexagonally shaped $\rm{BaF_2}$ detectors 
($\sim$~12~$X_0$) 
covering polar angles between $4^\circ$ and 
$30^\circ$ and the complete $2\pi$ azimuthal angle.
In front of each $\rm{BaF_2}$ module a 5 mm thick plastic scintillator 
was mounted for the identification of charged particles.
The resulting geometrical solid angle coverage of the combined system was $ 99 \%$ of $4\pi$.
The $\rm{BaF_2}$ crystals were read out by
photomultipliers providing a fast trigger, the CsI(Tl) crystals via photodiodes.
For further details see \cite{cb,taps,taps1}. 
\par
The invariant masses of the mesons were calculated from the measured 4-momenta of the 
decay photons. The
calibration of the Nb and $\rm{LH_2}$ data samples was carefully
cross checked by comparing the lineshapes for long-lived mesons, the 
$\pi^0$, $\eta$, and $\eta^\prime$. The decay lengths ($c\tau$) 
of 25.1~nm ($\pi^0$),
0.153~nm ($\eta$), and 0.001~nm ($\eta^\prime$) guarantee that these pseudoscalar
mesons will not decay inside the nucleus, hence the lineshapes
should not exhibit any difference for the two data samples.
Fig. \ref{cal_check} shows the comparison of the background subtracted invariant mass
distributions for $\pi^0\rightarrow\gamma\gamma$,
$\eta\rightarrow\pi^0\pi^0\pi^0\rightarrow6\gamma$ 
and $\eta^\prime\rightarrow\pi^0\pi^0\eta\rightarrow6\gamma$. 
Indeed, a difference in the lineshapes is not observed.
However, when comparing the $\omega\rightarrow\pi^0\gamma$ invariant mass distributions, 
we find a significant change in the lineshapes.
The left panel of Fig.~\ref{om_mom} shows the $\pi^0\gamma$ invariant mass distribution 
without further
cuts except for a three momentum cutoff of $|\vec{p}_{\omega}| < \rm{500~MeV/c}$.
The dominant background source is two pion production where one of the four photons
escapes the detection.
This probability was determined by Monte Carlo simulations
to be $14 \%$.
The resulting three photon final state is not distinguishable from the 
$\omega \rightarrow \pi^0 \gamma$ invariant mass.
The central panel of Fig.~\ref{om_mom} 
shows the invariant mass distribution obtained after 
background subtraction.
We observe the expected superposition of decays outside of the nucleus at the nominal vacuum mass 
with decays occuring inside the nucleus, 
responsible for the shoulder towards lower invariant masses.
The high mass part of the $\omega$ mass signal appears identical for the Nb and $\rm{LH_2}$ 
targets, indicating that this part is dominated by $\omega$ meson decays in vacuum. 
These decays are eliminated by matching the right hand part of the Nb invariant mass 
spectrum to the $\rm{LH_2}$ data (see central panel of Fig.~\ref{om_mom})
and by subtracting the two spectra from each other. 
For this normalization the integral of the undistorted spectrum corresponds to $75 \%$ of the 
counts in the Nb spectrum. This is in good agreement with  
a theoretical prediction obtained from a transport code calculation \cite{muehl,muehl1}.
There, about $16 \%$ of the total decays are predicted to occur inside the nuclear medium 
($\rho > 0.1\cdot \rho_0$) without any FSI and $3 \%$ of the events are distorted
due to FSI in the mass range of $0.6 ~\rm{GeV/c^2} < M_{\pi^0 \gamma} < 0.9~\rm{GeV/c^2}$. 
In addition, 
$9 \%$ of the events are moved towards lower masses due to the $\Delta$ decay kinematics. 
The right panel of Fig.~\ref{om_mom} shows the resulting in-medium signal
along with a Voigt fit (Breit-Wigner folded with Gaussian) to the data.
We obtain an  $\omega$ in-medium mass of
$\rm{M_{medium}=[722_{-2}^{+2}(stat)_{-5}^{+35}(syst)]}~\rm{MeV/c^2}$.
This corresponds to a lowering of the $\omega$-mass by $8~\%$ with respect to the vacuum value at an
estimated average nuclear density of $0.6~\rho_0$ in line with the assumptions in \cite{messch}.
Consistency with a scaling of the $\omega$-mass
by $m = m_0(1-0.14 \rho / \rho_0)$ 
is found \cite{brown}. 
Within this scenario the width is governed by the experimental
resolution of $\Gamma = 55~\rm{MeV/c^2}$ (FWHM).
The systematic uncertainty mainly reflects different assumptions for the
subtraction of decays of the $\omega$ mesons in vacuum. The fraction of these
decays was varied within a broad range from $80 \%$ to $45 \%$ (the central and right panel of Fig. 
\ref{om_mom} correspond to $75 \%$). The case with $45 \%$ corresponds to the upper bound of
the systematic uncertainty (+35 MeV). This extreme scenario would,
however, require an increase of the in-medium width of the $\omega$ by almost
an order of magnitude. 
\begin{figure}
  \vspace*{-0.9cm}
 \includegraphics[width=.83\columnwidth]{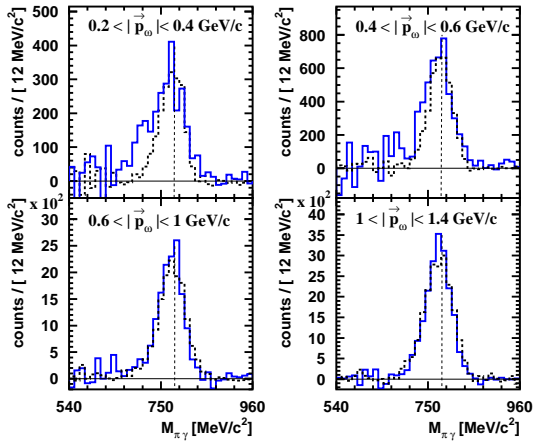}\hspace*{-0.1cm}
  \vspace*{-0.37cm}
  \caption[omega inviariant mass]{(color online). $\pi^0 \gamma$ mass spectrum after background 
subtraction and FSI suppression ($T_{\pi^0} > 150$ ~MeV) for different $\omega$ momentum bins.
Solid histogram: Nb data, dashed histogram: $\rm{LH_2}$ data. 
\label{om_dep}}
  \vspace*{-0.54cm}
\end{figure}
\par
\begin{figure}
  \includegraphics[width=.73\columnwidth]{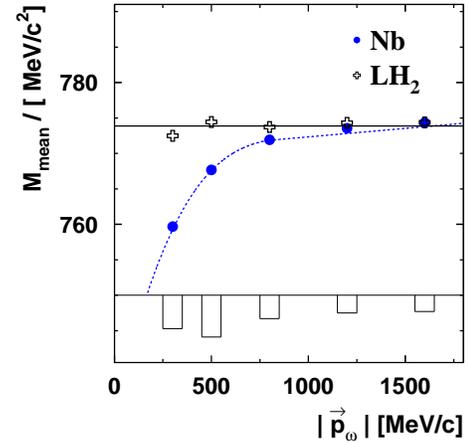}
  \vspace*{-0.2cm}
  \caption[mom dep]{(color online). Mean value of the $\pi^0 \gamma$ invariant mass 
as a function of
the $\omega$ momentum at an estimated average density of $0.6~\rho_0$ for the Nb data 
(circles) and
the $\rm{LH_2}$ (crosses) along with a fit.
The systematic errors are determined by varying the integration range 
of $\rm{0.65~\rm{GeV/c^2}<M_{\pi^0 \gamma}<0.9~GeV/c^2}$ by $\rm{50~MeV/c^2}$ and are
 shown as a bar chart
for the Nb data.
In contrast to the peak positions the mean values do not reach the nominal vacuum mass 
of $\rm{782~MeV/c^2}$ due to the detector response function.
\label{om_dep1}}
\end{figure}

Furthermore, the dependence of the signal on the $\omega$ momentum has been studied.
It is expected that only low-momentum $\omega$ mesons 
(with a corresponding low velocity) 
decay inside the nucleus and carry information on the in-medium properties of
the $\omega$ meson. Fig. \ref{om_dep} shows the $\pi^0 \gamma$ invariant mass distribution  
after background 
subtraction and  FSI suppression ($T_{\pi^0} > 150$ ~MeV) for different
$\omega$-momentum bins.
A pronounced modification of the 
lineshape is only observed for $\omega$ momenta in the range of 
$\rm{200~MeV/c}< |$~$\vec{p_\omega}$ $|<\rm{400~MeV/c}$.
In Fig. \ref{om_dep1} the mean value of the mass distribution is plotted 
against the three momentum for the $\rm{LH_2}$ and the Nb data.
This result might allow to extract the momentum dependence of the
$\omega$-nucleus potential \cite{muehl,muehl1}.
\par
In summary, we have investigated the in-medium modifications of $\omega$ mesons in
photoproduction experiments using the Crystal Barrel/TAPS detector at the ELSA 
accelerator facility in Bonn. 
When comparing data from a 
$\rm{LH_2}$ target with data taken with a Nb target, we find a pronounced modification of 
the $\omega$ meson mass in the nuclear medium for 
$\omega$ mesons with momenta less than 500~MeV/c.
The in-medium mass has been determined to  
$\rm{M_{medium}=[722_{-2}^{+2}(stat)_{-5}^{+35}(syst)]}~\rm{MeV/c^2}$
at an estimated average nuclear density of $0.6~\rho_0$.
The width is found to be $\Gamma = 55~\rm{MeV/c^2}$
and is dominated  by the experimental resolution.
The momentum dependence of the signal shows that 
only low-momentum $\omega$ mesons contribute to the downward 
mass shift.
In contrast, $\omega$ mesons with high momenta decay outside the nucleus, 
exhibiting an invariant mass distribution corresponding to $\omega$ decays in vacuum.  
First evidence for a lowering of the $\omega$ mass in the nuclear medium has
been observed.
\par
We gratefully acknowledge stimulating discussions with W. Cassing, S. Leupold,
U. Mosel, and in particular with P. M\"uhlich. We thank the accelerator group of ELSA as well as
the technicians and scientists of the HISKP in Bonn, the PI in Bonn and the
II. Physikalisches Institut in Giessen.
This work was supported by the Deutsche Forschungsgemeinschaft, SFB/Transregio
16, and the Schweizerischer Nationalfond. U. Thoma thanks for an Emmy-Noether grant from the DFG.

\end{document}